\documentclass[
reprint,
aps,pra,
superscriptaddress,
twocolumn,
longbibliography,
secnumarabic,
footinbib]{revtex4-1}
\usepackage[T1]{fontenc}
\usepackage[latin9]{inputenc}
\usepackage{hyperref}
\usepackage{array}
\usepackage{multirow}
\usepackage{mathrsfs}
\usepackage{amsmath}
\usepackage{amsthm}
\usepackage{amssymb}
\usepackage{dcolumn}
\usepackage{bm}
\usepackage{graphicx}
\usepackage{subcaption}
\usepackage{diagbox}

\makeatletter
\newtheorem{thm}{Theorem}
\newtheorem{lem}[thm]{Lemma}

\newtheorem{cor}[thm]{Corollary}
\makeatother

\begin{document}
\title{Unified Approach to Witness Nonentanglement-Breaking Quantum Channels}

\author{Yi-Zheng Zhen}
\email{zhenyz@sustech.edu.cn}
%\thanks{These authors contributed equally to this work.}
\affiliation{Institute for Quantum Science and Engineering and Department of Physics, Southern University of Science and Technology, Shenzhen, Guangdong 518055, P.~R.~China}
\affiliation{Hefei National Laboratory for Physical Sciences at Microscale and Department of Modern Physics, University of Science and Technology of China, Hefei, Anhui 230026, P.~R.~China}

\author{Yingqiu Mao}
%\thanks{These authors contributed equally to this work.}
%\email{myingqiu@mail.ustc.edu.cn}
\affiliation{Hefei National Laboratory for Physical Sciences at Microscale and Department of Modern Physics, University of Science and Technology of China, Hefei, Anhui 230026, P.~R.~China}
\affiliation{CAS Center for Excellence and Synergetic Innovation Center in Quantum Information and Quantum Physics, University of Science and Technology of China, Hefei, Anhui 230026, P.~R.~China}

\author{Kai Chen}
%\email{kaichen@ustc.edu.cn}
\affiliation{Hefei National Laboratory for Physical Sciences at Microscale and Department of Modern Physics, University of Science and Technology of China, Hefei, Anhui 230026, P.~R.~China}
\affiliation{CAS Center for Excellence and Synergetic Innovation Center in Quantum Information and Quantum Physics, University of Science and Technology of China, Hefei, Anhui 230026, P.~R.~China}

\author{Francesco Buscemi}
%\email{buscemi@i.nagoya-u.ac.jp}
\affiliation{Graduate School of Informatics, Nagoya University, Chikusa-ku, 464-8601 Nagoya, Japan}

\author{Oscar Dahlsten}
\email{dahlsten@sustech.edu.cn}
\affiliation{Institute for Quantum Science and Engineering and Department of Physics, Southern University of Science and Technology, Shenzhen, Guangdong 518055, P.~R.~China}
\affiliation{London Institute for Mathematical Sciences, 35a South Street Mayfair, London W1K 2XF, United Kingdom}
\affiliation{Wolfson College, University of Oxford, Linton Road, Oxford OX2 6UD, United Kingdom}

%\date{\today}

\begin{abstract}
The ability of quantum devices to preserve or distribute entanglement is essential in employing quantum technologies. 
Such ability is described and guaranteed by the nonentanglement-breaking (nonEB) feature of participating quantum channels.
For quantum information applications relying on entanglement, the certification of the nonEB feature is indispensable in designing, testing, and benchmarking quantum devices. 
Here, we develop a simple and direct approach for the certification of nonEB quantum channels. 
By utilizing the prepare-and-measure test, we derive a necessary and sufficient condition for witnessing the nonEB channels, which is applicable in almost all experimental scenarios.
The approach not only unifies and simplifies existing methods in the standard scenario and the measurement-device-independent scenario, but also further the nonEB channel certification in the semi-device-independent scenario.
\end{abstract}

\maketitle

\section{Introduction}

\begin{figure}[b!]
\centering
\includegraphics[width=.9\columnwidth]{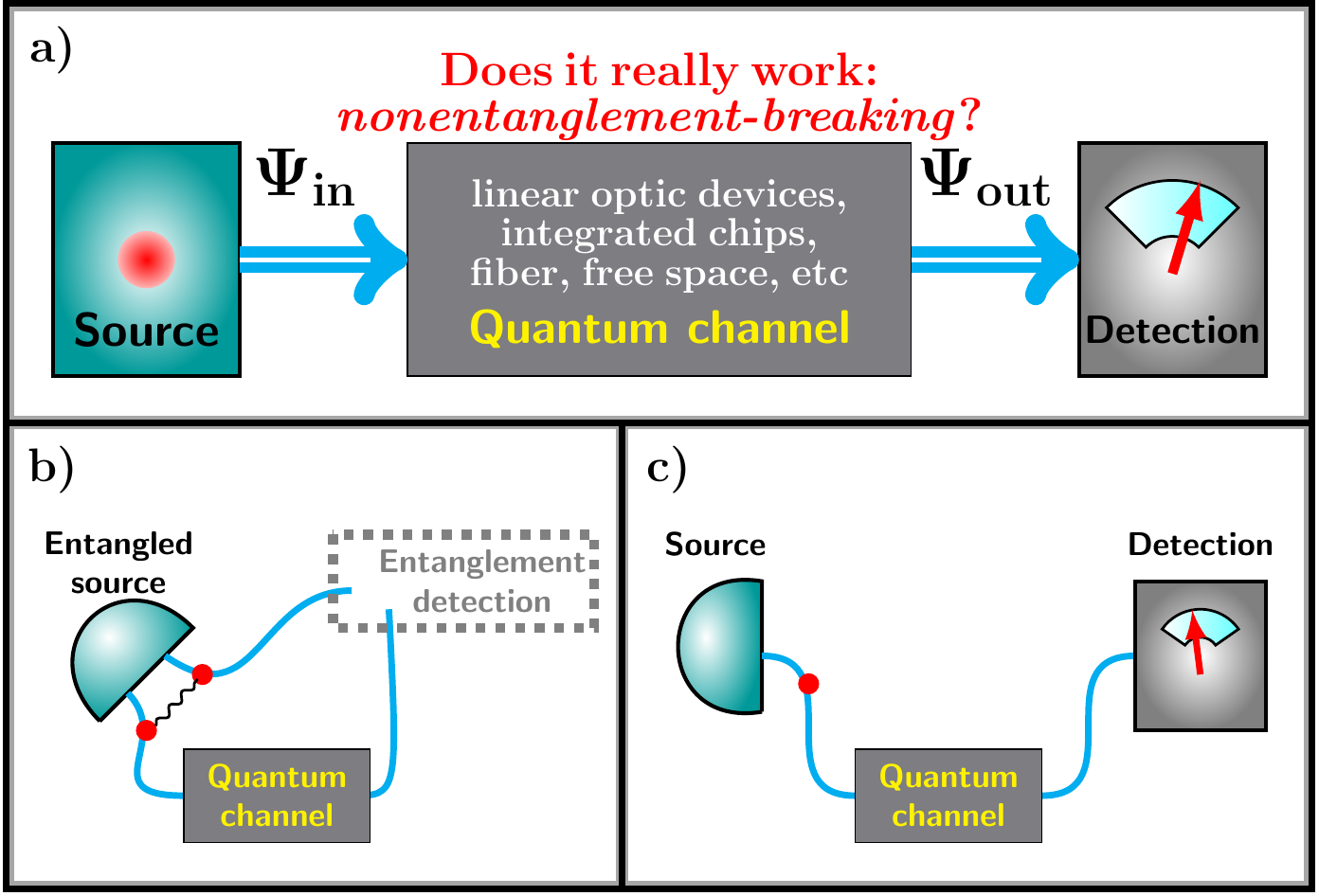}
\caption{\label{fig:concept}
a) The implementation of many quantum information tasks requires the nonEB quantum channel.
b) The test of quantum channels using entangled sources. 
c) The prepare-and-measure test without using
entangled sources.
In this work we show a unified and efficient method to study prepare-and-measure methods in various experimental scenarios.}
\end{figure}

Quantum entanglement \cite{Einstein1935pr, Schrodinger1935, Schrodinger1936} is of great value in the application of quantum information technologies \cite{Curty2004prl, jozsa2003role}.
Verifying the maintenance of quantum entanglement of realistic devices is thus important for performing quantum information tasks \cite{Galindo2002rmp, Braunstein2005rmp, Scarani2009rmp, Caruso2014rmp}.
Such devices generally transmit or store quantum states and are described by the concept of quantum channels.
To test whether these devices can preserve entanglement is equivalent to verify the non-entanglement-breaking (nonEB)
\cite{Horodecki2003rmp} feature of corresponding quantum channels.
Therefore, the certification of the nonEB feature of an unknown quantum channel
is crucial for guaranteeing the functionality of quantum devices (see Fig.~\ref{fig:concept}a).

Various methods can be applied to certify nonEB quantum channels. 
A natural method is using entangled sources (see Fig.~\ref{fig:concept}b).
By sending one subsystem of an entangled state through the channel, the entanglement detection \cite{Horodecki2009rmp, Guhne2009pr, Brunner2014rmp, Friis2019nrp, Buscemi2012prl, Branciard2013prl, Cavalcanti2013pra, Xu2014prl, Nawareg2015sr, Verbanis2016prl, Goh2016njp, Zhen2016pra, Yuan2016pra, Rosset2018pra, Bowles2018prl, Wiseman2007prl, Reid2009rmp, Cavalcanti2009pra, Cavalcanti2016rpp, Zhen2019e} at the output side can be used to infer the nonEB feature of the tested channel.
To optimize the certification, the maximally entangled state is usually required. Thus, the application of this method is restricted by the quality of entangled sources in practice.  
Another technological difficulty that can be involved is about correlated problem in the entanglement detection, e.g., the long-distance entanglement distribution.

To reduce experimental difficulty and costs, the prepare-and-measure (P\&M) methods \cite{Chuang1997pra, Mohseni2008pra, Piani2015josab, Pusey2015josab, DallArno2017prsa, Hoban2018njp, Agresti2019qst, Rosset2018prx, Yuan2019, Uola2019, Uola2019a, Guerini2019, Mao2019, Graffitti2019, Schmid2019, Zhang2019} can be adopted  (see Fig.~\ref{fig:concept}c).
By sending single-copy quantum states into the channel and measuring the output states directly, the input-output correlation reveals the nonEB feature of the tested channel.
In this sense, the P\&M methods do not require entangled sources and in principle can certify nonEB channels in the simplest way.

Existing P\&M methods, e.g., the process-tomography method \cite{Chuang1997pra, Mohseni2008pra}, channel steering \cite{Piani2015josab, Pusey2015josab}, semiquantum signalling games \cite{Rosset2018prx}, input-output games \cite{Yuan2019, Uola2019, Uola2019a, Guerini2019}, apply to different experimental situations.
For instance, tomography and input-output games characterize quantum channels based on the accurate preparations and measurements; channel steering and semiquantum signalling games are immune to detection-side imperfections but rely on accurate preparation of quantum states.
Because these methods detect nonEB feature from different perspectives, it is also hard to conclude to what extent a given input-output correlation can tolerate imperfections from experimental instruments.
These motivate the investigation of a general and unified P\&M nonEB detection approach.

In this paper, we formulate a unified and efficient P\&M approach to detect nonEB channels. 
The approach can be applied in almost all experimental scenarios considering trustworthiness of experimental instruments. 
For the general P\&M test on quantum channels, we derive a necessary and sufficient condition that a nonEB channel can be certified.
Based on this condition, the nonEB feature is detected via the violation of an inequality, whereas different bounds corresponds to different experimental scenarios.
Particularly, the approach can detect nonEB channels when only the dimension 
of quantum states are assumed in experiments.
Our results not only reduces experimental cost of nonEB channel tests in various experimental scenario, but also can be used to inspect the least requirements to exhibit the nonEB feature of a device.

\section{P\&M tests on nonEB channels}

The quantum channel is a completely positive and trace-preserving map ${\cal N}^{B\leftarrow A}$, which maps an arbitrary quantum state $\rho_{A}$ of system $A$ to a quantum state $\rho^{B}={\cal N}^{B\leftarrow A}(\rho^{A})$ of system $B$.
A quantum channel is nonEB if and only if it cannot be described by an entanglement-breaking (EB) channel in the following form \cite{Horodecki2003rmp},
\begin{equation}
{\cal N}_{{\rm EB}}^{B\leftarrow A}\left(\rho^{A}\right)=\sum_{k}{\rm tr}\left[E_{k}^{A}\rho^{A}\right]\tau_{k}^{B}.\label{eq:EB-channel}
\end{equation}
Here, $E_{k}^{A}$ are POVM elements satisfying $0\leqslant E_k \leqslant \mathbb{I}$ and $\sum_{k}E_{k}=\mathbb{I}$, and $\tau_{k}^{B}$ are quantum states.

\begin{figure}[h!]
\centering
\includegraphics[width=.9\columnwidth]{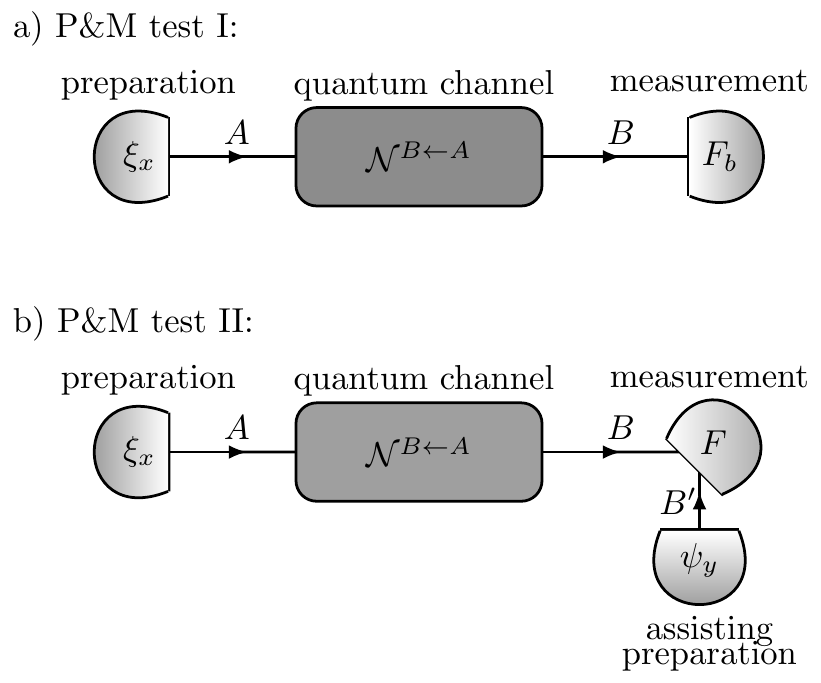}
\caption{\label{fig:prepare-measure-test}
Two types of P\&M tests on unknown quantum channels. 
a) The P\&M test I: the state $\xi_{x}$ is randomly prepared and $F_{b}$ denotes the POVM element associated with outcome $b$.
b) The P\&M test II: states $\xi_{x}$ and states $\psi_{y}$ are randomly prepared, and the measurement $F$ is fixed.}
\end{figure}

The EB channel is equivalent to a measure-and-prepare process, i.e., the process of measuring input state $\rho$ on a POVM $\{E_{k}\} $ and then producing another state $\tau_{k}$ according to the outcome $k$.
Consequently, for any entangled state $\rho_{{\rm ent}}^{A^{\prime}A}$ with one subsystem $A$ transmitting through an EB channel ${\cal N}_{{\rm EB}}^{B\leftarrow A}$, the output state $\rho^{A^{\prime}B}=({\rm id}^{A^{\prime}}\otimes{\cal N}_{{\rm EB}}^{B\leftarrow A})\rho_{{\rm ent}}^{A^{\prime}A}$ must be separable.
To detect the nonEB channel without entangled sources, in this work we focus on two kinds of P\&M tests, termed P\&M test I and P\&M test II (see Fig.~\ref{fig:prepare-measure-test}).

In the P\&M test I, a quantum state $\xi_{x}$, labelled by $x$, is randomly prepared and sent into an unknown channel ${\cal N}$. 
The output state ${\cal N}(\xi_{x})$ is then measured and an outcome $b$ is obtained.
% (for simplicity, here we put labels of measurement settings and outcomes together).
Denote the POVM element associated with $b$ as $F_{b}$. The probability to obtain $b$ given the input label $x$ is
\begin{equation}
P_{\cal N}^{\rm I}\left(b|x\right)={\rm tr}\left[{\cal N}^{B\leftarrow A}\left(\xi_{x}^{A}\right)F_{b}^{B}\right].
\label{eq:prob-prepa-measu}
\end{equation}
If the measurement $F_{b}$ is replaced with a fixed measurement assisted by another random state $\psi_{y}$, labelled by $y$, we have the P\&M test II.
Denote the POVM element associated with this outcome as $F$. 
The probability to obtain this outcome given states labels $x$ and $y$ is
\begin{equation}
P_{\cal N}^{\rm II}\left(x,y\right)={\rm tr}\left[{\cal N}^{B\leftarrow A}\left(\xi_{x}^{A}\right)\otimes\psi_{y}^{B^{\prime}}F^{BB^{\prime}}\right].\label{eq:prob-prepa-measu-assist}
\end{equation}

Both tests do not require entangled states. The P\&M test I is more suitable for testing distribution channels that transmitting quantum states to a remote place.
The P\&M test II is more suitable for testing memories that storing quantum states at a certain place.
In experiments of P\&M tests, based on the statistics $P_{\cal N}^{\rm I(II)}$, nonEB channels can be detected using the following theorem.
\begin{thm}\label{thm:exist-nEB-witness}
In P\&M tests I and II, the statistics of EB
channels always satisfy
\begin{eqnarray}
{\cal W}_{\cal N}^{\rm I(II)} & = &
\boldsymbol{w}\cdot\boldsymbol{P}_{\cal N}^{\rm I(II)}\geqslant C_{\rm EB}^{\rm I(II)},\label{eq:inequality}\\
C_{\rm EB}^{\rm I(II)} & = & \min_{{\cal N}_{{\rm EB}}}{\cal W}_{{\cal N}_{\rm EB}}^{\rm I(II)},\label{eq:CEB}
\end{eqnarray}
respectively, where $\boldsymbol{w}$ is a set of real coefficients.

A nonEB channel ${\cal N}$ can be certified in a P\&M test if and only if the inequality is violated.
\end{thm}

To prove Theorem \ref{thm:exist-nEB-witness}, we recall that, based on the Choi\textendash Jamio\l{}kowski isomorphism \cite{Jamiokowski1972rmp, Choi1975laa}, the EB feature of $\cal N$ is fully characterized by the entanglement of the Choi state:
\begin{equation}
\sigma_{{\cal N}}^{A^{\prime}B}=\left({\rm id}^{A^{\prime}}\otimes{\cal N}^{B\leftarrow A}\right)\left(\Phi_{+}^{A^{\prime}A}\right),\label{eq:Choi-state}
\end{equation}
where $\Phi_{+}=\sum_{mn}\left|mm\right\rangle \left\langle nn\right|/d$ is the maximally entangled state. We have the following Lemma \cite{Horodecki2003rmp}. 
\begin{lem}[Horodecki-Shor-Ruskai]
\label{lem:EB-condition}
The Choi state $\sigma_{{\cal N}}^{A^{\prime}B}$ of an EB channel ${\cal N}^{B\leftarrow A}$ is a separable density matrix satisfying ${\rm tr}_{B}[\sigma_{{\cal N}}^{A^{\prime}B}]=\mathbb{I}^{A^{\prime}}/d_{A^{\prime}}$. 
\end{lem}
Therefore, by using perfect entangled source, any nonEB channel $\cal N$ can be certified by producing the Choi state $\sigma_{\cal N}$  and performing a suitable entanglement witness \cite{Horodecki2009rmp, Guhne2009pr}.
Even without entangled source, such entanglement witness method can be extended to the P\&M approach.

\begin{proof}[Proof of Theorem 1]
Let $\mathbf{P}_{\cal N}$ be the collection of statistics $P_{\cal N}^{\rm I}(b|x)$ or $P_{\cal N}^{\rm II}(x,y)$, where the states and measurements can be unknown.
The set of $\mathbf{P}_{{\cal N}_{{\rm EB}}}$
for all EB channels ${\cal N}_{{\rm EB}}$ is a convex set under the convex combination.
Denote this set as $\mathfrak{C}_{\rm EB}$.
To see this, consider two statistics $\mathbf{P}_{{\cal N}_{\rm EB}^{(1)}}$ and $\mathbf{P}_{{\cal N}_{\rm EB}^{(2)}}$ produced by EB channels ${\cal N}_{\rm EB}^{(1)}$ and ${\cal N}_{\rm EB}^{(2)}$, respectively.
The convex combination of both, i.e., $\mathbf{P}_{{\cal N}_{q}}=q\mathbf{P}_{{\cal N}_{\rm EB}^{(1)}}+(1-q)\mathbf{P}_{{\cal N}_{\rm EB}^{(2)}}$ for $0\leqslant q\leqslant1$, can always be produced by another EB channel ${\cal N}_{\rm EB}^{(3)}$.
This is guaranteed by the definition of the EB channel.
Let ${\cal N}_{\rm EB}^{(1,2)}(\rho)=\sum_k{\rm tr}[E_k^{(1,2)}\rho]\tau_k^{(1,2)}$, where $\sum_k E_k^{(1,2)}=\mathbb{I}$ and $\tau_{k}^{(1,2)}$ are quantum states.
Then, ${\cal N}_q$ is equivalent to ${\cal N}_{\rm EB}^{(3)}(\rho)=\sum_k {\rm tr}[E_k^{(3)}\rho]\tau_k^{(3)}$ with $\{E_k^{(3)}\}=\{qE_k^{(1)},\dots;(1-q)E_k^{(2)},\dots\}$ and $\{\tau_k^{(3)}\}=\{\tau_k^{(1)},\dots;\tau_k^{(2)},\dots\}$.
It can be verified that ${\cal N}_{\rm EB}^{(3)}$ is a well-defined EB channel since $\sum_k E_k^{(3)}=\mathbb{I}$ and $P_{{\cal N}_{\rm EB}^{(3)}}=P_{{\cal N}_q}$.
Therefore, $\mathfrak{C}_{EB}$ is convex.

Based on the hyperplane separation theorem, two disjoint convex sets, e.g. $\mathfrak{C}_{\rm EB}$ and $\{\mathbf{P}_{\cal N}\} $ with $\mathbf{P}_{\cal N}\notin\mathfrak{C}_{\rm EB}$, can be distinguished by a linear inequality.
This inequality in general has the form of ${\cal W}_{\cal N}=\boldsymbol{w}\cdot\mathbf{P}_{\cal N}$, where $\boldsymbol{w}$ is a set of real parameters.
The bound $C_{\rm EB}$ for all EB channels is then the minimal value of ${\cal W}_{{\cal N}_{\rm EB}}$.
The violation of this bound implies that the tested channel is nonEB.
\end{proof}

%We give the proof, as well as proofs for corollaries in corresponding subsections in Supplemental Material.

%Since neither the forms of states or measurements nor the dimensions of quantum states were specified, this 
Theorem \ref{thm:exist-nEB-witness} actually gives a unified approach that can be applied in all experimental scenarios. 
%The EB channel bound $C_{\rm EB}$ is in a general form here, but 
When a detailed experimental condition is considered, the bound $C_{\rm EB}$ will have a clear and analytical form, such that the set of probabilities produced by nonEB channels is disjoint with the set of probabilities produced by the EB channels. 
For convenience of the following discussions, we introduce two operators 
\begin{subequations}\label{eq:W}
\begin{eqnarray}
W_{\rm I} & = & \sum_{x,b}w_{xb}\xi_{x}^{\rm T}\otimes F_{b},\label{eq:W-I}\\
W_{\rm II} & = & \sum_{x,y}w_{xy}\xi_{x}^{\rm T}\otimes\psi_{y}^{\rm T}.\label{eq:W-II}
\end{eqnarray}
\end{subequations}
Here, the superscript $\rm T$ denotes the transpose.

\subsection{The device-dependent scenario}

In the standard or device-dependent (DD) scenario, all experimental instruments can be assumed trusted or controlled well.
The desired state preparations and measurements can be realized perfectly. 
In this case, P\&M test II is equivalent to P\&M test I,
%Particularly, when the measurement in P\&M test II is the projective measurement onto the maximally entangled state, i.e. $F=\Phi_+=\sum_{mn}|mm\rangle\langle nn|/d$.
which can be verified from the fact that a general measurements is equivalent to a measurement on the state with an ancilla \cite{Nielsen2010}.
The EB bounds for P\&M test I and II are
\begin{subequations}
\begin{eqnarray}
C_{\rm EB}^{\rm I, DD} & = & d_{A}\min_{\sigma}{\rm tr}\left[W_{\rm I}\sigma\right],\label{eq:CEB-DD-I}\\
C_{\rm EB}^{\rm II, DD} & = & \frac{d_{A}}{d_{B}}\min_{\sigma}{\rm tr}\left[W_{{\rm II}}\sigma\right],\label{eq:CEB-DD-II}
\end{eqnarray}
\end{subequations}
where $\sigma$ is a separable state satisfying ${\rm tr}_{B}[\sigma^{AB}]=\mathbb{I}^A/d_A$.
Here, the fixed measurement in P\&M test II is the projective measurement onto the maximally entangled state, i.e. $\Phi_+=\sum_{mn}|mm\rangle\langle nn|/d$.
The violation of above bounds, i.e., ${\cal W}_{\cal N}^{\rm I(II)}<C_{{\rm EB}}^{\rm I(II),DD}$, implies the nonEB feature of the tested channel ${\cal N}$.

Particularly, by using P\&M test I (or II) and properly selecting $\{\boldsymbol{w},\xi_x,F_b\}$  (or $\{\boldsymbol{w},\xi_x,\psi_y,F\}$), any nonEB channel can be detected with a negative inequality value.
Similar to the entanglement witness \cite{Horodecki2003rmp, Guhne2009pr} for entangled states, Theorem \ref{thm:exist-nEB-witness} provides a witness for any nonEB channel.
\begin{cor}[NonEB channel witness]
\label{cor:dd-nEB-witness}
For any nonEB channel ${\cal N}_{{\rm nEB}}$, there always exists a P\&M test I (II) such that ${\cal W}_{{\cal N}_{\rm nEB}}^{\rm I(II)}<0$ whereas $C_{\rm EB}^{\rm I(II),DD}=0$.
\end{cor}

The proof is placed in Appendix \ref{app:proof}, where we use the entanglement witness of the associated Choi state to give the form of $\{\xi_x,F_b\}$ (or $\{\xi_x,\psi_y,F\}$) in the P\&M test I (or II).

%Because experimental instruments can be controlled perfectly i
In this scenario, the quantum process tomography can be applied to characterize unknown quantum channels.
To obtain the process matrix of the channel, experimental resources are usually consuming because of the large number of preparation and measurement settings
\cite{Mohseni2008pra}.
Instead of obtaining full information of the channel, the nonEB feature is detected with less state preparations and measurements with Corollary \ref{cor:dd-nEB-witness}.
Precisely, for a quantum system with dimension $d$, the tomography method typically involves a number of $d^{4}$ preparation and measurement settings; while in the witness this number can be reduced to $d^{2}$.

\subsection{The measurement-device-independent scenario}

The witness method in the DD scenario is based on the precise realization of desired measurements, which in practice is difficult to guarantee. 
For the situation with adversaries, the user may also only have access to untrusted measurement devices.
An eavesdropper may control the detection efficiencies to always simulate a nonEB channel, at the same time steal transmitted quantum information without being detected \cite{Brassard2000prl}. 
To obtain strict security and perform faithful implementation of nonEB channel detection, the witness method should be improved to the measurement-device-independent (MDI) scenario.

The MDI scenario is also important because in practice the functionality of preparation instruments is much easier to be guaranteed than that of measurement instruments.
In this scenario, state preparations are assumed to be perfect, while measurements are completely unknown. 
The EB bounds for two tests can be proven as
\begin{subequations}
\begin{eqnarray}
C_{\rm EB}^{\rm I, MDI} & = & d_{A}\min_{\sigma_{\rm  I},F_b}{\rm tr}\left[W_{\rm I}\sigma_{\rm I}\right],\label{eq:CEB-MDI-I}\\
C_{\rm EB}^{\rm II, MDI} & = & d_{A}d_{B}\min_{\sigma_{\rm II}}{\rm tr}\left[W_{\rm II}\sigma_{\rm II}\right],\label{eq:CEB-MDI-II}
\end{eqnarray}
\end{subequations}
respectively.
Here, $\sigma_{\rm I}$ and $\sigma_{\rm II}$ are separable states satisfying ${\rm tr}_{B}[\sigma_{\rm I}^{AB}]=\mathbb{I}^A/d_A$ and ${\rm tr}_{B}[\sigma_{\rm II}^{AB}]\leqslant\mathbb{I}^A/d_A$, respectively, and the violation of above bounds implies the nonEB feature of the tested channel.

In fact, the P\&M test I reduces to channel steering in the MDI scenario; see \cite{Piani2015josab, Pusey2015josab} and the recent work \cite{Guerini2019}.
Since the untrusted measurement does not provides enough information to recognize all nonEB channels, the witness result in Corollary~\ref{cor:dd-nEB-witness} does not hold.
In contrast, the P\&M test II can be developed as a witness for nonEB channels in the MDI scenario.
This can be understood by the result that a trusted measurement can be equivalently performed by an untrusted measurement with a trusted source \cite{Buscemi2012prl, Buscemi2012cmp, Cavalcanti2013pra}.
\begin{cor}[MDI nonEB channel witness]\label{cor:MDI-nEB-witness}
For any nonEB channel ${\cal N}_{{\rm nEB}}$, there always exists an P\&M test II such that ${\cal W}_{{\cal N}_{\rm nEB}}^{{\rm II}}<0$ whereas $C_{\rm EB}^{{\rm II,MDI}}=0$.
\end{cor}

The proof is placed in Appendix \ref{app:proof}, where again we use the entanglement witness of the associated Choi state to give the form of $\{\xi_x,\psi_y\}$ for the P\&M test II.

The P\&M test II simplifies and improves the semiquantum signalling game \cite{Rosset2018prx} for detecting nonEB channels.
Using informationally complete sets of quantum states \cite{Nielsen2010, Wilde2017} as channel input, the semiquantum signalling game defines a partial order for all quantum channels \cite{Rosset2018prx}, where EB channels stand at the bottom.
Instead of exhibiting a partial order for all quantum channels, the MDI nonEB channel witness has a lower requirements on input states.
Thus, Corollary \ref{cor:MDI-nEB-witness} can be adopted in real experiments even when state preparations are not perfect.
%  has lower experimental requirements on preparation instruments.
%This property benefits the test with limited preparations.
%For instance, in time-bin-and-phase encoding photonic qubits, states in the bin basis can be prepared with much higher fidelity than the states of the phase bases \cite{Mao2019}.
%In this case, methods based on perfect preparations should be applied very carefully.
%By characterizing the imperfect preparation instruments, Theorem \ref{thm:exist-nEB-witness} is still efficient in the MDI scenario.

\subsection{The semi-device-independent scenario}

To further weaken assumption of experimental instruments, we consider the scenario where both preparations and measurements are untrusted.
%the MDI scenario, the quality of sources have to be considered.
In fact, if all experimental instruments are fully untrusted, then we enter the device-independent scenario and no nonEB channel can be certified.
This is because the statistics of any P\&M test on any channel can always be explained by an EB channel with a higher dimension \cite{DallArno2017prsa}.
%Therefore, utilizing entangled sources and performing the loophole-free Bell test is up-to-now the only method.

Fortunately, since the quantum system usually has a finite size, the dimension of the quantum system is usually bounded.
This motivates the detection of nonEB channels in the semi-device-independent (SDI) scenario.
The application of Theorem \ref{thm:exist-nEB-witness} provides EB bounds as follows
\begin{subequations}\label{eq:CEB-SDI}
\begin{eqnarray}
C_{\rm EB}^{\rm I, SDI} & = & d_{A}\min_{\sigma_{\rm  I},\xi_x,F_b}{\rm tr}\left[W_{\rm I}\sigma_{\rm I}\right],
\label{eq:CEB-SDI-I}\\
C_{\rm EB}^{\rm II, SDI} & = & d_{A}d_{B}\min_{\sigma_{\rm II},\xi_x,\psi_y}{\rm tr}\left[W_{\rm II}\sigma_{\rm II}\right].
\label{eq:CEB-SDI-II}
\end{eqnarray}
\end{subequations}
Here, $\sigma_{\rm I}$ and $\sigma_{\rm II}$ are separable states satisfying ${\rm tr}_{B}[\sigma_{\rm I}^{AB}]=\mathbb{I}^A/d_A$ and ${\rm tr}_{B}[\sigma_{\rm II}^{AB}]\leqslant\mathbb{I}^A/d_A$, respectively, and the minimization is also taken over all states or POVMs in their associate Hilbert spaces with maximal dimension $d_A$ or $d_B$.
The violation of these bounds implies the nonEB feature of the tested channel.
\begin{cor}\label{cor:sdi}
In the SDI scenario, when dimensions of input states and output states (assisting states) are $d_A$ and $d_B$, respectively, a quantum channel $\cal N$ is certified as nonEB if there exists a P\&M test I (II) such that ${\cal W}_{\cal N}^{\rm I(II)}<C_{\rm EB}^{\rm I(II),SDI}$.
\end{cor}

The proof is placed in Appendix \ref{app:proof}.
Here, we straightforwardly extend the P\&M tests I and II to the SDI scenario. 
Due to limited information on preparation and measurement, the witness result may not hold the SDI scenario, i.e. there may be nonEB channels escaping the certification.
Despite this, Corollary \ref{cor:sdi} is still efficient if one considers low dimensions or large number of states and measurements, as we will show in the following example.

\section{Example: the depolarizing channel}

To show the properties of the P\&M tests in various experimental scenarios, let us consider the certification of the depolarizing channel,
\begin{equation}
{\cal N}_{\gamma}\left(\rho\right)=\gamma\rho+\left(1-\gamma\right)\frac{\mathbb{I}}{d},\label{eq:depolarzing-channel}
\end{equation}
where $0\leqslant\gamma\leqslant1$ and $d$ is dimension.
%The Choi state of ${\cal N}_{\gamma}$ is the isotropic state $\sigma_{{\cal N}_{\gamma}}=\gamma\Phi_{+}+(1-\gamma)\mathbb{I}/d^{2}$.
%Based on Lemma~\ref{lem:EB-condition}, the entanglement region $\gamma>1/(d+1)$ \cite{Werner1989pra} is also the nonEB region of ${\cal N}_{\gamma}$. 
It can be analytically calculated that the nonEB region is $\gamma>1/(d+1)$.
We leave the qudit case and detailed derivations in Appendix \ref{app:dep-cha}, and mainly discuss the qubit case for simplicity.
Let states and POVM elements chosen from eigenstates of three Pauli matrices $\sigma_x$, $\sigma_y$, and $\sigma_z$, which are denoted as $|+/-\rangle$, $|R/L\rangle$, and $|0/1\rangle$, respectively. Here, $|\pm\rangle=(|0\rangle\pm|1\rangle)/\sqrt{2}$ and $|R/L\rangle=(|0\rangle\pm i|1\rangle)/\sqrt{2}$.

In the DD scenario, the application of Corollary \ref{cor:dd-nEB-witness} can be realized by considering the entanglement witness of the associated Choi state.
An efficient P\&M test I can be designed as randomly inputting eigenstates of $\sigma_i$ into the channel and measuring the output states on the same $\sigma_i$.
With $w_{++}=w_{--}=w_{RR}=w_{LL}=-1/2$,  $w_{+-}=w_{-+}=w_{RL}=w_{LR}=1/2$, and $w_{01}=w_{10}=1$ (other $w$'s are 0), we have $C_{\rm EB}^{\rm I,DD}=0$, and ${\cal W}_\gamma^{\rm I}<0$ implies $\gamma>1/3$ exactly.
However, if measurement $\sigma_i$ is imperfect, a false certification may occur.
For example, suppose the actual measurements are $\tilde{\sigma}_{i}=\epsilon_{i}\sigma_{i}$, where $\epsilon_{i}\in(0,1]$ are detection efficiencies. %dependent on measurement settings.
A direct application of Corollary~\ref{cor:dd-nEB-witness} gives $\gamma>\epsilon_{z}/\sum_{i}\epsilon_{i}$.
If detection efficiencies satisfy $\epsilon_{z}<(\epsilon_{x}+\epsilon_{y})/2$, the
EB depolarizing channels in the region $\epsilon_{z}/\sum_{i=x,y,z}\epsilon_{i}<\gamma\leqslant1/3$ would be falsely certified.

To avoid this problem, the nonEB channel certification can be applied in the MDI scenario. 
By again considering the entanglement witness of the Choi state, an efficient P\&M test II can be designed.
Let $\xi_x$ and $\psi_y$ be randomly prepared in the same basis $\sigma_i$. 
If the untrusted measurement faithfully implements $\Phi_{+}$, with the same $\boldsymbol{w}$, we also have $C_{\rm EB}^{\rm II,DD}=0$ and ${\cal W}_\gamma^{\rm II}<0$ implies $\gamma>1/3$.
%The same result can also be obtained in the DD scenario.
Even if the measurement $\Phi_{+}$ is inaccurate, no EB channel can pass the test.
To see this, suppose the actual POVM element is $\Phi^\prime=\varepsilon |\phi_{\theta}\rangle\langle\phi_{\theta}|$, where $\varepsilon$ is the efficiency and $|\phi_{\theta}\rangle =\cos\theta|00\rangle +\sin\theta|11\rangle $.
A direct application of Corollary~\ref{cor:MDI-nEB-witness} gives $\gamma>\gamma_\theta=1/(1+2\sin2\theta)$.
For $\theta\in(-\pi/12,7\pi/12)$, the nonEB region $1/3\leqslant\gamma_\theta<\gamma\leqslant1$ can be certified. 
For other values of $\theta$, nonEB channel would be certified and one has to change to another inequality. 
Therefore, imperfect measurements can only weaken the performance of the MDI witness but never cause false certification.

Less quantum states can also certifies nonEB channels with our method.
%When only eigenstates of $\sigma_{z}$ and $\sigma_{x}$ can be prepared in P\&M test II.
Using $\xi_x,\psi_y\in\{|+/-\rangle,|0/1\rangle\}$ and $w_{01}=w_{10}=1$ and $w_{+-}=w{-+}=-w_{++}==-w_{--}=1/2$, the EB bound is calculated as $0$ and the violation gives $1/2<\gamma\leqslant1$.
If the input states are further reduced to $\xi_x\in\{|0\rangle\langle 0|,|+\rangle\langle +|\}$ and $\psi_y\in\{|1\rangle\langle 1|,|-\rangle\langle -|\}$, with $w_{01}=w_{+-}=-w_{0-}=1$, the EB bound is still $0$ and the violation certifies the same $1/2<\gamma\leqslant 1$. 
There is a gap between certified region and theoretical nonEB region because less states are used.

In the SDI scenario, due to the analytical difficulty in calculating EB bounds, we numerically discuss the minimal $\gamma$ that can be certified for fixed dimensions $d_A$ and $d_B$.
%For P\&M test I, we consider the statistics produced by randomly inputting eigenstates of Pauli matrices into the depolarizing channel and randomly measuring the output states with Pauli matrices.
%For P\&M test II, we consider the statistics produced by preparing $\xi_{x}$ and $\psi_{y}$ randomly as eigenstates of Pauli matrices and implementing the measurement as $\Phi_{+}$.
%To simplify calculations, we use $w_{xb(xy)}=\pm1$  and the PPT criterion \cite{Horodecki2009rmp, Guhne2009pr} to generate separable states $\sigma_{\rm I(II)}$.
As shown in Table \ref{tab:sdi-depol}, our method can still certify nonEB channels efficiently.
Particularly, when $d_{A}=d_{B}=2$, almost all nonEB depolarizing channels can be detected.
When $d_{A}$ and $d_{B}$ increase, it becomes hard for the weak nonEB depolarizing channel to pass the test.
In P\&M test I, the nonEB feature is certified only when $d_B$ is small, while in P\&M test II, the same minimal $\gamma$ is certified for the same pair of $d_{A}$ and $d_{B}$ because of the symmetry between $\xi_x$ and $\psi_y$ in this case.
When $d_A$ and $d_B$ are large, due to the specific inequalities and limited states and measurements in the tests, the nonEB feature is not revealed.
This can be improved with a better optimization or using more quantum states.
%Despite this, the results show the practicality of the inequality method.

\begin{table}[h!]

\begin{subtable}{.5\columnwidth}
\centering
\begin{tabular}{|c|c|c|c|c|}
\hline 
\diagbox{$d_A$}{$d_B$} & 2 & 3 & 4 & 5\tabularnewline
\hline 
2 & 0.34 & 0.39 & 0.58 & 0.58\tabularnewline
\hline 
3 & 0.58 & 0.67 & - & -\tabularnewline
\hline 
4 & 0.70 & 0.75 & - & -\tabularnewline
\hline 
5 & 0.82 & 0.87 & - & -\tabularnewline
\hline 
\end{tabular}
\caption{The P\&M test I}
\end{subtable}%
\begin{subtable}{.5\columnwidth}
\begin{tabular}{|c|c|c|c|c|}
\hline 
\diagbox{$d_A$}{$d_B$} & 2 & 3 & 4 & 5\tabularnewline
\hline 
2 & 0.34 & 0.58 & 0.70 & 0.82\tabularnewline
\hline 
3 & 0.58 & - & - & -\tabularnewline
\hline 
4 & 0.70 & - & - & -\tabularnewline
\hline 
5 & 0.82 & - & - & -\tabularnewline
\hline 
\end{tabular}
\caption{The P\&M test II}
\end{subtable}
\caption{\label{tab:sdi-depol}
The minimal $\gamma$ certified by the P\&M tests on the qubit depolarizing channel ${\cal N}_{\gamma}$, given dimensions of Hilbert spaces as $d_{A}$ and $d_{B}$, respectively.}
\end{table}

Our results can be used to determine the minimal experimental requirements to reveal the nonEB feature.
Consider the qubit depolarizing channel with $\gamma=0.55$.
The above discussion shows that both P\&M tests can certify this channel in the DD and MDI scenarios.
Particularly, in the MDI scenario one can use only four states.
This channel can also be certified even when both preparation and measurement instruments are untrusted but have fixed dimensions, precisely, $(d_A=2,d_B\leqslant3)$ in P\&M test I or $d_A=d_B=2$ in the P\&M test II.

\section{Conclusion}

In this paper, we have formulated a unified framework for the P\&M test on the nonEB channel.
We have derived a necessary and sufficient condition for certifying a nonEB quantum channel, then applied it to various experimental scenarios for two kinds of P\&M tests. 
In the DD scenario, because accurate and faithful state preparations and measurements can be performed, the nonEB channel witness can be directly realized.
However, such certification is not reliable when measurement instruments are imperfect. 
We then applied the inequality criterion in the MDI scenario, and show that P\&M test II can be formulated as a witness.
The certification in the MDI scenario is not only robust to imperfect measurements, but also applicable for relaxed requirements of state preparations.
Considering real-life trustworthiness of sources, we further extend the inequality method to the SDI scenario.
Based on dimensions of Hilbert spaces solely, both SDI P\&M tests certify nonEB channels effectively.

The inequality criterion uses different EB bounds in associated scenarios for the same inequality. 
These bounds have clear and compact forms, most of which can be calculated analytically.
After a P\&M test, based on the violation of different EB bounds in corresponding scenarios, the minimal experimental requirements on exhibiting the nonEB feature can be obtained.
Our results complement the entanglement detection in the temporal situation via the certification of nonEB quantum channels, and can be adopted in the evaluation and designation of real quantum devices.

\begin{acknowledgements}
We thank Kavan Modi, T. Kraft, and R. Uola for valuable discussions. 
Y.-Z. Z. especially thanks Prof. Qinghe Mao for numerous advices and encouragements throughout writing this paper.
This work has been
supported by the National Natural Science Foundation of China (Grants No.~U1801661 and No.~11575174).
F.B. acknowledges support from the Japan Society for the Promotion of Science (JSPS) KAKENHI, Grant No. 19H04066.

Y.-Z. Z. and Y. M. contributed equally to this work.
\end{acknowledgements}

\appendix

\section{\label{app:proof}Proofs for Corollaries}

Here, we give the proofs of Corollary \ref{cor:dd-nEB-witness}, \ref{cor:MDI-nEB-witness}, and \ref{cor:sdi}.

For the P\&M test I, the input-output correlation is
\begin{align}
P_{{\cal N}}^{{\rm I}}\left(b|x\right) & ={\rm tr}\left[{\cal N}\left(\xi_{x}\right)F_{b}\right]={\rm tr}\left[\xi_{x}{\cal N}^{\dagger}\left(F_{b}\right)\right]\nonumber\\
 & =d_{A}{\rm tr}\left[\xi_{x}^{{\rm T}}\otimes{\cal N}^{\dagger}\left(F_{b}\right)\Phi_{+}\right]\nonumber\\
 & =d_{A}{\rm tr}\left[\left(\xi_{x}^{{\rm T}}\otimes F_{b}\right){\rm id}\otimes{\cal N}\left(\Phi_{+}\right)\right]\nonumber\\
 & =d_{A}{\rm tr}\left[\left(\xi_{x}^{{\rm T}}\otimes F_{b}\right)\sigma_{{\cal N}}\right],\label{eq:p-I}
\end{align}
where ${\rm tr}[{\cal N}(A)B]={\rm tr}[A{\cal N}^{\dagger}(B)]$ and ${\rm tr}[A^{\rm T}B]=d{\rm tr}[A\otimes B\Phi_{+}]$ have been used, and $d_{A}$ is the dimension of ${\cal H}_{A}$.
Then, the inequality expression is
\begin{equation}
{\cal W}_{\cal N}^{\rm I} =d_{A}{\rm tr}\left[W_{\rm I}\sigma_{\cal N}\right],
\end{equation}
with $W_{\rm I}=\sum_{x,b}w_{xb}\xi_{x}^{\rm T}\otimes F_{b}$.

In the DD scenario, both $\xi_x$ and $F_b$ are known.
Based on Lemma~\ref{lem:EB-condition}, the EB channel bound is
\begin{eqnarray}
C_{\rm EB}^{\rm I,DD} & = & d_{A}\min_{\sigma_{{\cal N}_{\rm EB}}}{\rm tr}\left[W_{\rm I}\sigma_{{\cal N}_{EB}}\right]\\
 & = & d_{A}\min_{\sigma_{{\rm sep}}}{\rm tr}\left[W_{\rm I}\sigma_{\rm sep}\right],\label{eq:ceb-dd-I-p}
\end{eqnarray}
where $\sigma_{\rm sep}$ is a separable state satisfying ${\rm tr}_{B}[\sigma_{\rm sep}^{AB}]=\mathbb{I}^{A}/d_{A}$. 
In the MDI scenario, $\xi_x$ is known while $F_b$ is unknown; while in the SDI scenario, both $\xi_x$ and $F_b$ are unknown except dimensions.
To exclude all effects from unknown terms, the EB channel bounds are
\begin{eqnarray}
C_{\rm EB}^{\rm I,MDI} & = & d_{A}\min_{\sigma_{\rm sep},F_b}{\rm tr}\left[W_{\rm I}\sigma_{\rm sep}\right],\label{eq:ceb-mdi-I-p}\\
C_{\rm EB}^{\rm I,SDI} & = & d_{A}\min_{\sigma_{\rm sep},\xi_x,F_b}{\rm tr}\left[W_{\rm I}\sigma_{\rm sep}\right]\label{eq:ceb-sdi-I-p},
\end{eqnarray}
where $\sigma_{\rm sep}$ satisfies the same condition as in Eq.~(\ref{eq:ceb-dd-I-p}), and the minimization is also taken over all $F_b$ or $\xi_x$ in associated operator spaces. 
%Thus, Eqs.~(\ref{eq:CEB-DD-I}), (\ref{eq:CEB-MDI-I}) and (\ref{eq:CEB-SDI-I}) are proved.

For the P\&M test II, the input-output correlation is 
\begin{eqnarray}
P_{{\cal N}}^{{\rm II}}\left(x,y\right) & =& {\rm tr}\left[{\cal N}\left(\xi_{x}\right)\otimes\psi_{y}F\right]\nonumber \\
 & =& {\rm tr}\left[\xi_{x}\otimes\psi_{y}\left({\cal N}^{\dagger}\otimes{\rm id}\right)F\right]\nonumber \\
 & =& d_{A}{\rm tr}\left[\left(\xi_{x}^{{\rm T}}\otimes\mathbb{I}\otimes\psi_{y}\right)\left(\mathbb{I}\otimes F\right)\left(\sigma_{{\cal N}}\otimes\mathbb{I}\right)\right]\nonumber \\
 & =& d_{A}d_{B}{\rm tr}\left[\xi_{x}^{{\rm T}}\otimes\psi_{y}^{{\rm T}}\tilde{\sigma}_{{\cal N}}\left(F\right)\right]\label{eq:p-II}
\end{eqnarray}
where% (Below is one step for the above equation
% & =d_{A}{\rm tr}\left[\left(\xi_{x}^{{\rm T}}\otimes\mathbb{I}\otimes\psi_{y}\right)\left(\mathbb{I}\otimes\left({\cal N}^{\dagger}\otimes{\rm id}\right)F\right)\left(\Phi_{+}\otimes\mathbb{I}\right)\right]\nonumber \\
\begin{equation}
\tilde{\sigma}_{{\cal N}}\left(F\right)=\frac{1}{d_{B}}{\rm tr}_{2}\left[\left(\mathbb{I}\otimes F\right)\left(\sigma_{{\cal N}}\otimes\mathbb{I}\right)\right]^{{\rm T}_{3}}.
\end{equation}
is an unnormalized state.
Here, ${\rm tr}_2$ acts on the second operator space, and ${\rm T}_3$ is the transpose on the third operator space. Then, the inequality expression is
\begin{equation}
{\cal W}_{\cal N}^{\rm II}  =d_{A}d_{B}{\rm tr}\left[W_{{\rm II}}\tilde{\sigma}_{{\cal N}}\left(F\right)\right],
\end{equation}
where $W_{{\rm II}}=\sum_{xy}w_{xy}\xi_{x}^{\rm T}\otimes\psi_{y}^{\rm T}$.

In the DD scenario, $\xi_x$ and $\psi_y$ can be prepared well. 
If we let the measurement be $F=\Phi_+$, then $\tilde{\sigma}_{\cal N}(\Phi_+)=\sigma_{\cal N}/d_B^2$.
Based on Lemma~\ref{lem:EB-condition}, the corresponding EB bound is
\begin{eqnarray}\label{eq:ceb-dd-II-p}
C_{\rm EB}^{\rm II,DD} & = & \frac{d_{A}}{d_{B}}\min_{\sigma_{{\cal N}_{\rm EB}}}{\rm tr}\left[W_{\rm II}\sigma_{{\cal N}_{\rm EB}}\right]\\
 & = & \frac{d_{A}}{d_{B}}\min_{\sigma_{\rm sep}}{\rm tr}\left[W_{{\rm II}}\sigma_{{\rm sep}}\right],
\end{eqnarray}
where $\sigma_{\rm sep}$ is a separable state satisfying ${\rm tr}_{B}[\sigma_{\rm sep}^{A^{\prime}B}]=\mathbb{I}^{A^{\prime}}/d_{A^{\prime}}$. 
In the MDI and SDI scenarios, the measurement $F$ is unknown, but for EB channels, the corresponding $\tilde{\sigma}_{\cal N}$ can be simplified.
The Choi state of an EB channel is separable, i.e., $\sigma_{{\cal N}_{\rm EB}}=\sum_{k}p_{k}\tau_{k}^\prime\otimes\tau_{k}$ with $\sum_{k}p_{k}\tau_{k}^\prime=\mathbb{I}^{A}/d_{A}$.
For any POVM element $F$,
\begin{eqnarray}
\tilde{\sigma}_{{\cal N}_{\rm EB}}\left(F\right) & = & \frac{1}{d_{B}}{\rm tr}_2\left[\left(\mathbb{I}\otimes F\right)\left(\sigma_{\cal N}\otimes\mathbb{I}\right)\right]^{{\rm T}_3}\nonumber\\
 & = & \sum_{k}p_{k}\tau_{k}^\prime\otimes{\rm tr}_2\left[F\left(\tau_{k}\otimes\frac{\mathbb{I}}{d_{B}}\right)\right]^{{\rm T}_3}\nonumber\\
 & = & \sum_{k}p_{k}\tau_{k}^\prime\otimes\tilde{\tau}_{k},
\end{eqnarray}
where $\tilde{\tau}_{k}$ is an unnormalized state satisfying ${\rm tr}[\tilde{\tau}_{k}]\leqslant$1. Thus, ${\rm tr}_{B}[\tilde{\sigma}_{{\cal N}_{\rm EB}}(F)]\leqslant\mathbb{I}^{A}/d_{A}$, and the EB channel bounds in the MDI and SDI scenarios are
\begin{eqnarray}
C_{\rm EB}^{\rm II,MDI} & = & d_{A}d_{B}\min_{\sigma_{\rm sep}}{\rm tr}\left[W_{\rm II}\sigma_{\rm sep}\right],\label{eq:ceb-mdi-II-p}\\
C_{\rm EB}^{\rm II,SDI} & = & d_{A}d_{B}\min_{\sigma_{\rm sep},\xi_x,\psi_y}{\rm tr}\left[W_{\rm II}\sigma_{\rm sep}\right],\label{eq:ceb-sdi-II-p}
\end{eqnarray}
where $\sigma_{\rm sep}$ satisfies the same condition as in Eq.~(\ref{eq:ceb-dd-II-p}).

\begin{proof}[Proof of Corollary 2-4]
Recall that for an arbitrary entangled state $\rho_{\rm ent}$, there always exists a witness $W$ such that ${\rm tr}[W\rho_{\rm ent}]<0$ while ${\rm tr}[W\rho_{\rm sep}]\geqslant0$ holds for all separable states $\rho_{\rm sep}$.
For conveninece, the witness $W$ can be decomposed as $W=\sum_{ij} c_{ij}A_{i}\otimes B_{j}$, where $c_{ij}$ are real coefficients and $A_{i}$ and $B_{j}$ are operators satisfying $0< A_i,B_j \leqslant \mathbb{I}$.

To prove Corollary 2 and 3, we can always choose $w_{xb}$ ($w_{xy}$), $\xi_{x}$ and $F_{b}$ ($\psi_y$) such that $W_{\rm I(II)}$ is an entanglement witness for the Choi state $\sigma_{{\cal N}_{\rm nEB}}$ of the tested nonEB channel ${\cal N}_{\rm nEB}$.
For the P\&M test I, let $w_{xb}=c_{xb}{\rm tr}[M_x]$, $\xi_x=M_x^{\rm T}/{\rm tr}[M_x]$, and $F_b=N_b$.
It can be verified that $W_{\rm I}=W$ and $C_{\rm EB}^{\rm I,DD}=0$. 
For the P\&M test II, similarly, let $w_{xy}=c_{xy}{\rm tr}[M_x]{\rm tr}[N_y]$, $\xi_x=M_x^{\rm T}/{\rm tr}[M_x]$, and $\psi_y=N_y^{\rm T}/{\rm tr}[N_y]$.
We also have $W_{\rm II}=W$.
Then, from Eqs.~(\ref{eq:ceb-dd-II-p}) and (\ref{eq:ceb-mdi-II-p}), we have the EB channel bounds $C_{\rm EB}^{\rm II,DD}=0$ and $C_{\rm EB}^{\rm II,MDI}=0$, respectively.
Therefore, Corollary~\ref{cor:dd-nEB-witness} and Corollary~\ref{cor:MDI-nEB-witness} are proved.

Corollary~5 naturally holds from Theorem~1 and (\ref{eq:ceb-sdi-II-p}).
\end{proof}

Notice that, in this proof if we use the singular value decomposition in writing the witness, we will have $W=\sum_{k=1}^{d}\tilde{A}_k\otimes \tilde{B}_k$, where $d$ is the minimal dimension of systems $A$ and $B$.
After transferring to $W_{\rm I(II)}$, the number of preparation and measurement settings, i.e., number of pairs $(\xi_x,F_b)$, is not more than $d^2$.

\section{\label{app:dep-cha}The depolarizing channel}

For the depolarizing channel ${\cal N}_{\gamma}$, the entanglement witness for its Choi state $\sigma_{{\cal N}_{\gamma}}$ is $W_{\rm dep}=\mathbb{I}/d-\Phi_{+}$ \cite{Werner1989pra}, based on which we can construct the nonEB channel witness in both DD and MDI scenarios.

In the DD scenario, we let the P\&M test I be the Table \ref{tab:depol-test-I}.
it can be verified that $W_{\rm I}=W_{\rm dep}$.
Thus, $C_{\rm EB}^{I,DD}=0$ and from Corollary~2, the negative value of 
\begin{equation}
{\cal W}_{{\cal N}_\gamma}^{\rm I}=\left(d-1\right)\left[1-\left(d+1\right)\gamma\right]<0
\end{equation}
implies the exactly nonEB region of depolarizing channel, i.e., $\gamma>1/\left(d+1\right)$.

\begin{table}[h]
\centering{}
\begin{tabular}{c|c|c|c}
$w=\left\{ w_{\mu}\right\} $ & $\Xi=\left\{ \xi_{\mu}\right\} $ & ${\cal F}=\left\{ F_{\mu}\right\} $ & $\mu$\tabularnewline
\hline 
\hline 
$w_{k}=1$ & $\xi_{k}=\left|k\right\rangle \left\langle k\right|$ & $F_{k}=\mathbb{I}-\left|k\right\rangle \left\langle k\right|$ & $k=1,\dots,d$\tabularnewline
\hline 
\multirow{2}{*}{$w_{\mu\nu}=-1/2$} & \multicolumn{2}{c|}{$(\xi_{\mu},F_{\nu})\in\{(+_{kl},+_{kl}),(-_{kl},-_{kl})$} & \multirow{4}{*}{$1\leqslant k<l\leqslant d$}\tabularnewline
 & \multicolumn{2}{c|}{$\phantom{(\xi_{\mu},F_{\nu})\in\{}(R_{kl},R_{kl}),(L_{kl},L_{kl})\}$} & \tabularnewline
\cline{1-3} \cline{2-3} \cline{3-3} 
\multirow{2}{*}{$w_{\mu\nu}=1/2$} & \multicolumn{2}{c|}{$(\xi_{\mu},F_{\nu})\in\{(+_{kl},-_{kl}),(-_{kl},+_{kl}),$} & \tabularnewline
\cline{2-3} \cline{3-3} 
 & \multicolumn{2}{c|}{$\phantom{(\xi_{\mu},F_{\nu})\in\{}(R_{kl},L_{kl}),(L_{kl},R_{kl})\}$} & \tabularnewline
\end{tabular}

\caption{\label{tab:depol-test-I}
A P\&M test I in the DD scenario.  
Here, $\{|k\rangle |k=1,\dots,d\}$ forms an orthonormal basis for the Hilbert space with dimension $d$.
The $\pm_{kl}$ and $R_{kl}/L_{kl}$ denote the operators projected onto the state $|\pm_{kl}\rangle =(|k\rangle \pm|l\rangle)/\sqrt{2}$ and $|R_{kl}/L_{kl}\rangle =(|k\rangle \pm i|l\rangle)/\sqrt{2}$, respectively.}
\end{table}

Suppose that in the experiment, the actual measurements are $\tilde{F}_{\mu}=\epsilon_{\mu}F_{\mu}$, where $\epsilon_{\mu}$ represents the detection efficiency.
For simplicity, we assume that the efficiency for measuring $|k\rangle\langle k|$ is $\epsilon_{z}$, for measuring $+_{kl}$ is $\epsilon_{x}$, and for measuring $-_{kl}$ is $\epsilon_{kl}=\epsilon_{y}$.
The inequality value is 
\begin{equation}
{\cal W}_{{\cal N}_\gamma}=\left[\epsilon_{z}-\gamma\epsilon_{z}-\gamma\frac{\epsilon_{x}+\epsilon_{y}}{2}d\right]\left(d-1\right).
\end{equation}
If we still apply ${\cal W}_{{\cal N}_\gamma}<0$, then we would obtain $\gamma>\gamma_{\epsilon}=2\epsilon_{z}/(2\epsilon_{z}+d\epsilon_{x}+d\epsilon_{y})$.
When $\epsilon_{z}\leqslant(\epsilon_{x}+\epsilon_{y})/2$, the EB region$\gamma_{\epsilon}<\gamma\leqslant1/(1+d)$ would be falsely certified as nonEB.
When $\epsilon_{z}>(\epsilon_{x}+\epsilon_{y})/2$,
the nonEB region $1/(d+1)<\gamma\leqslant\gamma_{\epsilon}$ would not be certified.

In the MDI scenario, the P\&M test II can be chosen as Table \ref{tab:depol-test-II}, we can also obtain $W_{\rm II}=W_{\rm dep}$.
The EB channel bound is $C_{\rm EB}^{\rm II,MDI}=0$ and  if the untrusted measurement implements $\Phi_+$ faithfully, then we have the inequality value
\begin{equation}
{\cal W}^{\rm II}_{{\cal N}_\gamma} = \frac{d-1}{d}\left[1-\left(d+1\right)\gamma\right].
\end{equation}
The violation also implies $\gamma>1/(d+1)$ such that all nonEB region is witnessed.

\begin{table}[h]
\centering
\begin{tabular}{c|c|c|c}
$w=\left\{ w_{\mu}\right\} $ & $\Xi=\left\{ \xi_{\mu}\right\} $ & $\Psi=\left\{ \psi_{\mu}\right\} $ & $\mu$\tabularnewline
\hline 
\hline 
$w_{kl}=1-\delta_{kl}$ & $\xi_{k}=\left|k\right\rangle \left\langle k\right|$ & $\psi_{l}=\left|l\right\rangle \left\langle l\right|$ & $k,l=1,\dots,d$\tabularnewline
\hline 
\multirow{2}{*}{$w_{\mu\nu}=-1/2$} & \multicolumn{2}{c|}{$(\xi_{\mu},\psi_{\nu})\in\{(+_{kl},+_{kl}),(-_{kl},-_{kl})$} & \multirow{4}{*}{$1\leqslant k<l\leqslant d$}\tabularnewline
 & \multicolumn{2}{c|}{$\phantom{(\xi_{\mu},F_{\nu})\in\{}(R_{kl},L_{kl}),(L_{kl},R_{kl})\}$} & \tabularnewline
\cline{1-3} \cline{2-3} \cline{3-3} 
\multirow{2}{*}{$w_{\mu\nu}=1/2$} & \multicolumn{2}{c|}{$(\xi_{\mu},\psi_{\nu})\in\{(+_{kl},-_{kl}),(-_{kl},+_{kl}),$} & \tabularnewline
\cline{2-3} \cline{3-3} 
 & \multicolumn{2}{c|}{$\phantom{(\xi_{\mu},F_{\nu})\in\{}(R_{kl},R_{kl}),(L_{kl},L_{kl})\}$} & \tabularnewline
\end{tabular}

\caption{\label{tab:depol-test-II}
A P\&M test II in the MDI scenario.  
Here, $\{|k\rangle |k=1,\dots,d\}$ forms an orthonormal basis for the Hilbert space with dimension $d$.
The $\pm_{kl}$ and $R_{kl}/L_{kl}$ denote the operators projected onto the state $|\pm_{kl}\rangle =(|k\rangle \pm|l\rangle)/\sqrt{2}$ and $|R_{kl}/L_{kl}\rangle =(|k\rangle \pm i|l\rangle)/\sqrt{2}$, respectively.}
\end{table}

To discuss the situation when $\Phi_+$ is not perfectly measured, we consider the qubit case and suppose the actual measurement is $\Phi^{\prime}=\varepsilon|\phi_{\theta}\rangle\langle \phi_{\theta}|$, where $\varepsilon\in\left(0,1\right]$ and $|\phi_\theta\rangle =\cos\theta|00\rangle +\sin\theta|11\rangle$.
The inequality value becomes
\begin{equation}
{\cal W}^{\rm II,MDI}_{{\cal N}_{\gamma}}=\frac{\varepsilon}{2}\left[1-\left(1+2\sin2\theta\right)\gamma\right].
\end{equation}
The negative value implies $\gamma>\gamma_{\theta}=1/(1+2\sin2\theta)$.
For $\theta\in(-\pi/12,\pi/4)$, we would have $1/3\leqslant\gamma_{\theta}<1$ and the nonEB region $\gamma\in(\gamma_{\theta},1]$ can be certified.
For other values of $\theta$, we would have either $\gamma>1$ or $\gamma<-1$, i.e., errors in the measurement destroy the test.

For the certification in the SDI scenario, we consider the P\&M tests I and II as following.
Denote the eigenstates of Pauli matrices $\sigma_x$, $\sigma_y$ and $\sigma_z$ as $|\pm\rangle$, $|R/L\rangle$, and $|0/1\rangle$, respectively.
In the P\&M test I, we consider the statistics $\boldsymbol{P}_{{\cal N}_\gamma}^{\rm I} = \{P_{{\cal N}_\gamma}^{\rm I}(b|x)|x,b\in\{0,1;+,-;R,L\}\}$.
In the P\&M test II, we consider the statistics $\boldsymbol{P}_{{\cal N}_\gamma}^{\rm II} = \{P_{{\cal N}_\gamma}^{\rm II}(xy)|x,y\in\{0,1;+,-;R,L\}\}$ and the measurement realized as $\Phi_+$.
We further choose the inequality expression as $w_{xb},w_{xy}\in\{\pm1\}$, and simplify the calculation by computing the minimal distance between $\boldsymbol{P}_{{\cal N}_\gamma}^{\rm I(II)}$ and $\boldsymbol{P}_{{\cal N}_{\rm EB}}^{\rm I(II)}$ generated by EB channels, instead of calculating each inequality individually.
That is, we numerically calculate
\begin{eqnarray}
{\rm Dist}_\gamma^{\rm I} &=& \min_{{\cal N}_{\rm EB},\xi_x,F_b} \left|\boldsymbol{P}_{{\cal N}_\gamma}^{\rm I}-\boldsymbol{P}_{{\cal N}_{\rm EB}}^{\rm I}\right|\nonumber\\
&\propto& \min_{\sigma_{\rm I},\xi_x,F_b} \sum_{x,b}\left|{\rm tr}\left[\xi_x^{\rm T}\otimes F_b\left(\sigma_{{\cal N}_\gamma}-\sigma_{\rm I}\right)\right]\right|,\\
{\rm Dist}_\gamma^{\rm II} &=& \min_{{\cal N}_{\rm EB},\xi_x,F_b(\psi_y)} \left|\boldsymbol{P}_{{\cal N}_\gamma}^{\rm II}-\boldsymbol{P}_{{\cal N}_{\rm EB}}^{\rm II}\right|\nonumber\\
&\propto& \min_{\sigma_{\rm II},\xi_x,F_b} \sum_{x,y}\left|{\rm tr}\left[\xi_x^{\rm T}\otimes \psi_y^{\rm T}\left(\sigma_{{\cal N}_\gamma}-\sigma_{\rm II}\right)\right]\right|,
\end{eqnarray}
for P\&M tests I and II, respectively.
The minimal $\gamma$ such that ${\rm Dist}^{\rm I(II)}_\gamma>0$ are concluded for each $d_A$ and $d_B$.

\bibliography{RefnonEB}

\end{document}